\documentclass[aps,prb,showpacs,twocolumn,floatfix,letterpaper,superscriptaddress,citeautoscript,10pt]{revtex4-1}

\usepackage[intlimits,sumlimits]{amsmath}
\usepackage{amsfonts,amssymb}
\usepackage{bm}
\usepackage{graphicx}
\usepackage{hyperref}

\hypersetup{%
pdftitle={Interaction corrections to the polarization function of graphene},%
pdfauthor={Inti Sodemann, Michael M. Fogler},%
pdfpagemode={UseNone},%
pdfstartview={FitH},%
breaklinks=true}

\begin{document}
\title{Interaction corrections to the polarization function of graphene}

\author{I. Sodemann}
\affiliation{Kavli Institute for Theoretical Physics, Santa Barbara, California 93106}
\affiliation{Department of Physics, University of Texas at Austin, Austin, Texas 78712}
\author{M. M. Fogler}
\affiliation{Kavli Institute for Theoretical Physics, Santa Barbara, California 93106}
\affiliation{University of California San Diego, 9500 Gilman Drive, La Jolla, California 92093}

\date{\today}

\begin{abstract}

The first-order interaction correction to the irreducible polarization function of pristine graphene is studied at arbitrary relation between momentum and frequency.
The results are used to calculate the dielectric function and the dynamical conductivity of graphene beyond the standard random-phase approximation.
The computed static dielectric constant compares favorably with recent experiments.

\end{abstract}

\pacs{
73.22.Pr, 
78.67.Wj, 
73.80.Vp  
}
\maketitle

\section{Introduction}
\label{sec:Introduction}

The influence of Coulomb interactions on electron properties of graphene is a subject of much interest~\cite{CastroNeto2009tep, Kotov2012eei}.
Prominent examples of interaction effects are the enhancement of the quasiparticle velocity $v$
above the typically quoted value $v = 1.0 \times 10^8\,\text{cm}/\text{s}$ at low momenta~\cite{Elias2011dcr, Li2008dcd, Siegel2011mbi} (as predicted
by the early theoretical work\cite{Gonzalez1994nfl, Gonzalez1999mfl}),
and the emergence of additional dispersion branches\cite{Bostwick2010oop, Walter2011esa} believed to be plasmarons --- bound states of electrons and plasmons.
The strength of Coulomb interactions in graphene is determined by the dimensionless parameter ($\hbar \equiv 1$ in this paper)
\begin{equation}
\alpha = \frac{e^2}{\kappa v} = \frac{2.2}{\kappa}\,,
\label{eqn:alpha}
\end{equation}
which can be controlled by varying the effective dielectric constant $\kappa$ of the graphene environment.

Interactions modify not only the quasiparticle properties but also the response functions of graphene, e.g.,
the irreducible (or proper) polarization $P(q,\omega)$.
This quantitiy is defined to be the coefficient of proportionality between the change in electron density and the self-consistently determined screened scalar potential.
The observables directly related to $P$ are the dielectric function $\epsilon$ and the longitudinal conductivity $\sigma$:
\begin{align}
\epsilon(q, \omega) &= 1 - \frac{2\pi e^2}{q} \,  P(q, \omega)\,,
\label{eqn:epsilon}\\
\sigma(q, \omega) &= e^2\, \frac{i \omega}{q^2}\, P(q, \omega)\,.
\label{eqn:sigma}
\end{align}
Our work is motivated in part by two recent experiments\cite{Reed2010tef, Wang2012mdq}
that suggested that the standard random phase approximation\cite{Mahan1990mpp} (RPA)
significantly underestimates the static dielectric function $\epsilon(q, 0)$ of graphene.
The RPA amounts to replacing the exact polarization function $P$ by the noninteracting value $P_0$.
For neutral graphene at zero temperature,
which is the case studied here, function $P_0$ is given by\cite{Gonzalez1994nfl}
\begin{equation}
P_0(q,\omega) = -\frac{q^2}{4 \sqrt{v^2 q^2 - (\omega + i0)^2}}
\label{eqn:P_0}
\end{equation}
at small enough $q$ and $\omega$ where the Dirac approximation is valid.
Equations~\eqref{eqn:epsilon} and \eqref{eqn:P_0} entail~\cite{Ando2002dca}
\begin{equation}
\epsilon_{\text{RPA}}(q, 0) = 1 + \frac{\pi}{2}\, \alpha\,.
\label{eqn:epsilon_RPA}
\end{equation}
For $\kappa = 1$ this formula gives $\epsilon_{\text{RPA}} \approx 4.6$.
On the other hand,
a much larger value $\epsilon = 15.4^{+39.6}_{-6.4}$
was inferred from inelastic x-ray scattering on bulk graphite\cite{Reed2010tef}.
Similarly, for graphene on a boron nitride substrate ($\kappa \approx 2.5$,
$\alpha \approx 0.9$) $\epsilon_{\text{RPA}} \approx 2.4$,
whereas the study of charge profile near Coulomb impurities by means of scanning tunneling microscopy suggests\cite{Wang2012mdq} $\epsilon = 3.0 \pm 0.1$.

\begin{figure}[t]
	\begin{center}
		\includegraphics[width=2.6in]{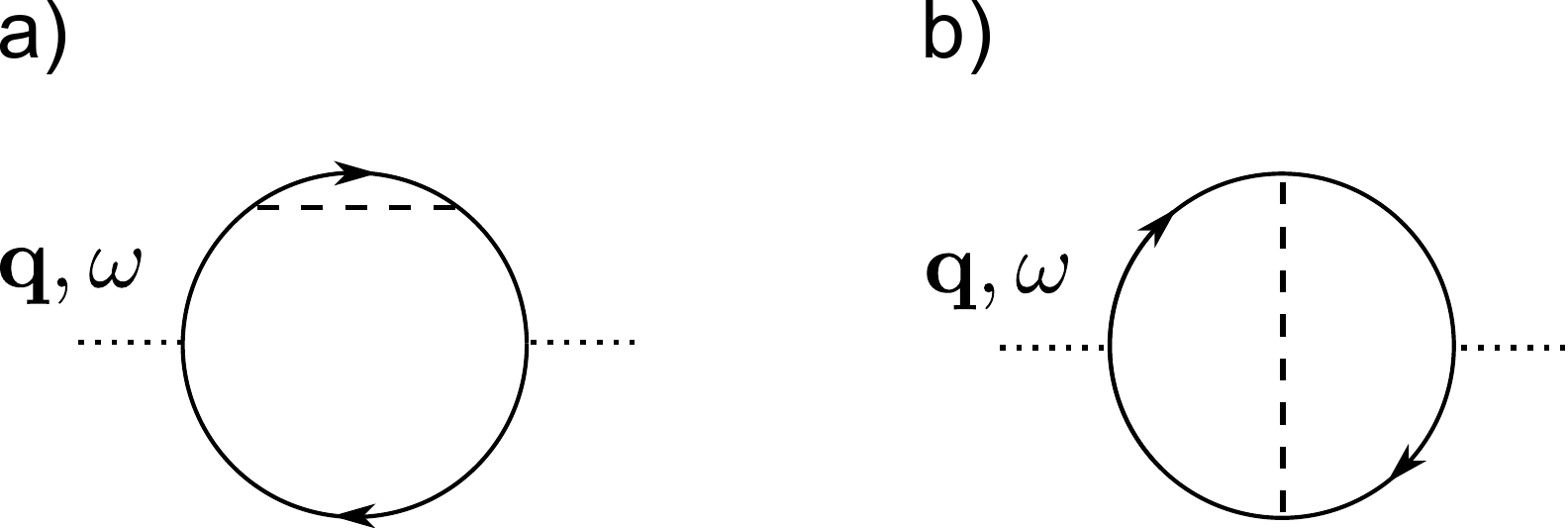}
	\end{center}
	\caption{First-order diagrams for the polarization function include the self-energy part (a) and the vertex part (b). Solid lines represent electron Green's function, dashed lines the Coulomb interaction, and dotted lines the external momentum and frequency.}
	\label{fig:Diagrams}
\end{figure}

The first-order interaction correction $P_1 = \mathcal{O}(\alpha)$ beyond the RPA is represented by the diagrams depicted in Fig.~\ref{fig:Diagrams}.
We show that by including this correction,
\begin{equation}
\label{eqn:pol}
P(q,\omega) \to P_0(q,\omega) + P_1(q,\omega)\,,
\end{equation}
one can significantly reduce the discrepancy between the theory and experiment.
Our result for $\epsilon(q, 0)$ is\footnote{Similar results have been obtained by F.~Guinea and also by A.~Principi and M.~Polini (private communications).}
\begin{equation}
\epsilon(q, 0) = 1 + \frac{\pi}{2}\, \alpha_q + 0.778 \alpha_q^2\,,
\quad \alpha_q \ll 1\,.
\label{eqn:epsilon_dc}
\end{equation}
(The difference between $\alpha_q$ and $\alpha$ will be clarified in Sec.~\ref{sec:results}.)
This formula yields $\epsilon \approx 3.0$ at $\alpha_q = 0.9$, in excellent agreement with Ref.~\onlinecite{Wang2012mdq}.
In turn, if we use Eq.~\eqref{eqn:epsilon_dc} for $\alpha_q = 2.2$, we get $\epsilon \approx 8.2$,
which is close to the lower estimate of the static dielectric constant in Ref.~\onlinecite{Reed2010tef}, although, the use of a
perturbative formula is questionable at such large interaction
strengths.

Another topic of interest is the dynamic response of graphene.
For example, in the ``optical'' limit $\omega\gg v q$,
Eqs.~\eqref{eqn:sigma} and \eqref{eqn:P_0} imply
that noninteracting Dirac fermions have a frequency-independent conductivity\cite{Ludwig1994iqh}
\begin{equation}
\sigma_0 (0,\omega)=  \frac{e^2}{4}\,.
\label{eqn:sigma_0}
\end{equation}
Accordingly, deviations of the infrared optical conductivity of graphene from the universal value $\sigma_0$ may signal interaction effects.
Experimentally, no such deviations have been observed\cite{Mak2008mot, Nair2008fsc, Li2008dcd}
while the theoretical calculation of the interaction corrections has been a subject of a debate.
Our analysis favors the result
\begin{equation}
\frac{\sigma(0, \omega)}{\sigma_0} - 1 \simeq \frac{19 - 6\pi}{12}\, \alpha\,,
\quad \alpha \ll 1\,,
\label{eqn:sigma_I}
\end{equation}
derived in~Refs.~\onlinecite{Mishchenko2008mci, Sheehy2009oto}.
We explain why a different coefficient was obtained in Ref.~\onlinecite{Juricic2010coi}.
The numerical smallness of $(19 - 6\pi) / 12 \approx 0.01$ may be the reason why
the interaction corrections have not been observed so far.

To get a more complete understanding of the interaction corrections
we also compute $P_1(q, \omega)$ at arbitrary $\omega / v q$ ratios.
This enables us to see the evolution of $P_1(q, \omega)$ in the full momentum-frequency parameter space, from the small value in the optical limit to the divergence at the spectral boundary $\omega = v q$ to a sizable effect in the static limit.
Besides theoretical interest,
motivation for this calculation comes from a rapidly burgeoning effort in probing
regimes of intermediate $\omega/v q$ by state-of-the-art experimental techniques, such as the near-field optics\cite{Fei2011ino, Fei2012gto, Chen2012oni} and electron energy loss spectroscopies\cite{Liu2010pps, Koch2010spp, Tegenkamp2011peh, Shin2011ooi}.
Such regimes are also pertinent for the electromagnetic response of graphene nanoribbons\cite{Ju2011gpf}.

The remainder of the paper is organized as follows. In Sec.~\ref{sec:results} we summarize our main results. Derivation of these results is outlined in Sec.~\ref{sec:derivation}. In Sec.~\ref{sec:discussion} we compare our findings with previous work. The Appendix~\ref{sec:optical} is devoted to the critical analysis of the controversy regarding the optical limit.


\section{Main results}
\label{sec:results}

In order to present our results we first need to explain the notation $\alpha_q$ in Eq.~\eqref{eqn:epsilon_dc} above.
Crudely, $v$ and $\alpha$ represent the bare velocity and the bare coupling constant of the theory,
whereas $v_q$ and $\alpha_q$ denote their renormalized values.
More precisely, $v_q$ is defined to be the phase velocity of quasiparticles with momentum $q$:
\begin{equation}
\label{eqn:vel_I}
v_q \equiv v + \frac{\Sigma_q}{q}\,,
\end{equation}
where $\Sigma_q$ is the on-shell self-energy.
At the level of the first-order perturbation theory one finds\cite{Gonzalez1994nfl, Gonzalez1999mfl}
\begin{equation}
\Sigma_q = \frac{e^2 q}{4\kappa}\, \ln \frac{\Lambda}{q}\,,
\label{eqn:Sigma_q}
\end{equation}
where $\Lambda$ is the high-momentum cutoff.
Therefore, $v_q$ and $\alpha_q$ are given by
\begin{align}
v_q &= v + \frac{e^2}{4\kappa}\,
\ln \frac{\Lambda}{q}\,,
\label{eqn:vel_II}\\
\alpha_q &= \frac{e^2}{\kappa v_q}
          = \left(\frac{1}{\alpha} + \frac{1}{4}\, \ln \frac{\Lambda}{q}\right)^{-1}\,.
\label{eqn:alpha_q}
\end{align}
According to the renormalization group approach,
such expressions are not just first-order approximations.
They are asymptotically exact at low enough $q$ where $\alpha_q \ll 1$.
However, $\alpha$ should be understood as the running coupling constant evaluated at some other cutoff $\Lambda$.
The choice of $\alpha$ is largely arbitrary because a change in $\alpha$ in Eq.~\eqref{eqn:alpha_q} can be absorbed into $\Lambda$.
Nevertheless, $\alpha_q$, which is determined by the observable quantity,
the phase velocity $v_q$, is unambiguous.
Relations between various observable quantities are expressible in terms of the renormalized parameters only. For example, phase velocities at two different momenta $q$ and $k$ are linked by the relation
\begin{equation}
\frac{v_q}{v_k} = 1 + \frac{\alpha_k}{4}\, \ln\frac{k}{q}\,,
\end{equation}
which is free from the arbitrary parameters $\alpha$ and $\Lambda$.
Similarly, the polarization function can and (if one desires higher accuracy) should
be expressed in terms of $\alpha_q$ and $v_q$. This point will be stressed again in Sec.~\ref{sec:discussion}.

\begin{figure*}
\begin{center}
\includegraphics[width=7.0in]{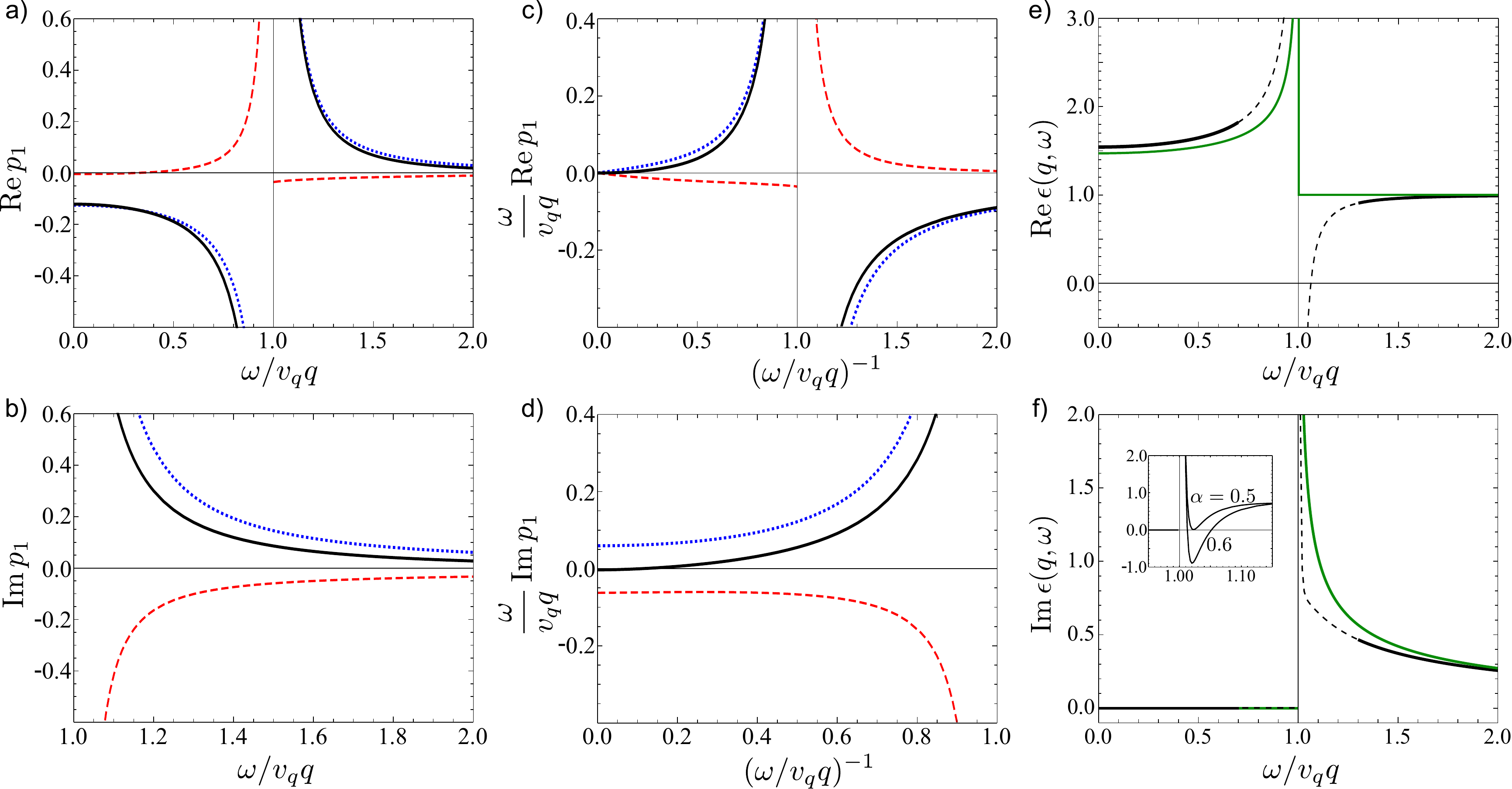}
\end{center}	
\caption{(Color online) Panels (a) and (b) depict function $p_1(x)$ where $x = \omega / v_q q$.
The red dashed line, the blue dotted line, and the black solid line are, respectively, the self-energy term, the vertex term,
and their sum, which is $p_1(x)$.
Although not shown in the figure,
very close to $x = 1$ the divergence of the self-energy term overwhelms that of the vertex term,
which causes the sign change of $\mathrm{Re}\, p_1(x)$ at
$x = 0.998$ and $\mathrm{Im}\, p_1(x)$ at $x = 1.005$.
Panels (c) and (d) illustrate $x p_1(x)$,
the quantity which (if multiplied by $4i$) gives the interaction correction to the universal conductivity,
$\sigma / \sigma_0 - 1$.
The meaning of the three curves is the same as in panels (a) and (b).
Panels (e) and (f) depict the real and imaginary
parts of the dielectric function for $\alpha_q = 0.3$.
The green solid line is the RPA result
and the black curves (solid and dashed) include the interaction correction.
The dashed part of the curve is within $\alpha_q$ from the absorption threshold.
The first-order perturbation theory is expected to be unreliable in that region.
Unphysical behavior near the threshold is exemplified by the negative sign of $\text{Im}\,\epsilon$ in the inset of panel (f). In panel (f) the second order interaction correction (black curve) becomes larger than the RPA value (green curve) at $\omega/v_q q \approx 9.1$ (not shown).
\label{fig:p1}
}
\end{figure*}

Let us now present our findings.
The first-order correction to the polarization function is written as
\begin{equation}
P_1(q,\omega) = \frac{\alpha_q q}{v_q}\, p_1(x)\,,
\label{eqn:p_1}
\end{equation}
where $x$ is the dimensionless ratio
\begin{equation}
x = \frac{\omega}{v_q q}
\label{eqn:x}
\end{equation}
and $p_1(x)$ is the complex dimensionless function, whose real and imaginary parts are displayed in Figs.~\ref{fig:p1}(a)--\ref{fig:p1}(d).
In general, $p_1(x)$ has to be evaluated numerically.
However,
analytical results are available in several limits,
as discussed later in this section.

The full polarization function to order $\mathcal{O}(\alpha_q)$ is
\begin{equation}
P(q,\omega)=\frac{q}{v_q}
\left[-\frac{1}{4}\frac{1}{\sqrt{1 - (x+i0)^2}} + \alpha_q p_1(x)\right]\,.
\label{eqn:P_from_p_1}
\end{equation}
The finite imaginary part of $P$ appears when $x$ exceeds unity, which we refer to as the absorption threshold.
Once $p_1(x)$ is known, one can use Eqs.~\eqref{eqn:epsilon} and \eqref{eqn:P_from_p_1} to get the dielectric function $\epsilon(q, \omega)$.
For example, in the static limit we obtain Eq.~\eqref{eqn:epsilon_dc}.
The real and imaginary parts of $\epsilon(q, \omega)$ as a function of $x$ are
illustrated by Figs.~\ref{fig:p1}(e) and~\ref{fig:p1}(f) for the case
of a suitably low $\alpha_q = 0.3$.
The RPA predictions are also plotted in Figs.~\ref{fig:p1}(e) and~\ref{fig:p1}(f) for comparison.
As one can see, the RPA underestimates $\mathrm{Re}\, \epsilon(q, \omega)$ at $x < 1$ and overestimates it at $x > 1$.
Near the absorption threshold, at $|1 - x| < \alpha_q$,
the first-order results are deemed unreliable and are shown by the dashed line.
In that region higher order corrections become as important as the first-order ones.
An example of inapplicability of the first-order perturbation theory near $x = 1$
is the negative sign of $\mathrm{Im}\, \epsilon(q, \omega)$ at sufficiently large $\alpha_q$,
see the inset of Fig.~\ref{fig:p1}(f).

Let us next discuss the analytical results for the function $P_1(q, \omega)$.
This function can be written as
\begin{equation}
P_1(q,\omega) = 2 P_a(q,\omega) + P_b(q,\omega)\,,
\end{equation}
where $P_a$ and $P_b$ are the self-energy and vertex terms represented by
the corresponding diagrams in Fig.~\ref{fig:Diagrams}.
(The factor of $2$ comes from the symmetry of the self-energy diagram.)
Below we describe these terms separately.

For the self-energy contribution we have the result
\begin{equation}
2 P_a(q,\omega)=\frac{\alpha q}{8 \pi v}\,
\left[\frac{\pi}{2}\frac{1}{(1-y^2)^{3/2}}\ln \frac{\Lambda}{q}+I_a(y)\right]\,,
\label{eqn:Pa2}
\end{equation}
where $y = \omega / (v q) + i 0$.
The expression for function $I_a(y)$, which is rather cumbersome, is given by Eq.~\eqref{eqn:I_a} (Sec.~\ref{sec:derivation}).

Equation~\eqref{eqn:Pa2} is written in terms of the ``bare'' parameters.
As explained above, we should rewrite it in terms of the renormalized ones.
To do so we combine $2 P_a(q, \omega)$ with the zeroth order polarization function $P_0(q, \omega)$ [Eq.~\eqref{eqn:P_0}] and get,
to the order $\mathcal{O}(\alpha_q)$,
\begin{equation}
\begin{split}
P_0(q,\omega) + 2 P_a(q,\omega)
&= -\frac{1}{4}\, \frac{q}{v_q}\, \frac{1}{\sqrt{1 - (x + i0)^2}}\\
&+ \frac{1}{8 \pi}\, \frac{\alpha_q q}{v_q}\, I_a(x + i0)\,.
\label{eqn:P02Pa}
\end{split}
\end{equation}
The desired renormalized $P_0(q,\omega)$ and $2 P_a(q,\omega)$ are, respectively,
the first and the second line of Eq.~\eqref{eqn:P02Pa}.
The limiting forms of such renormalized $2 P_a$ are as follows.
In the static limit we find
\begin{gather}
2 P_a(q,0) = -\frac{C_a}{4}\, \frac{\alpha_q q}{v_q}\,,
\label{eqn:Pa_dc}\\
C_a = \frac{1}{8} - \frac{\ln 2}{2} + \frac{1}{\pi}\left(G - \frac{1}{6}\right) \approx 0.017\,,
\label{eqn:C_a}
\end{gather}
%
where $G = 0.916\ldots$ is the Catalan constant.
In the optical limit, we obtain
\begin{equation}
2 P_a(q,\omega) \simeq \frac{1}{16}\,
\frac{\alpha_q q^2}{i \omega}\,,
\quad \omega \gg v_q q\,.
\label{eqn:P02Pa_optical}
\end{equation}
We expect that the accuracy of this expression is improved if in place of $\alpha_q$ one uses $\alpha_k$, where $k$ is such that $\omega = v_k k$.

Near the absorption threshold, $|x - 1| \ll 1$, we get
\begin{equation}
2 P_a \simeq
\frac{1}{16}\, \frac{\alpha_q q}{v_q}\, \left[
\frac{2\ln 2 - \frac{5}{6}}{(1 - x^2)^{3/2}}
-\frac{1}{3}\, \frac{1}{(1 - x^2)^{1/2}}
\right]\,.
\label{eqn:Pa_theshold}
\end{equation}
Note that the coefficient in front of the dominant $(1 - x^2)^{-3/2}$ singularity depends on the renormalization procedure.

Our result for the vertex term is
\begin{equation}
P_b(q, \omega) = -\frac{1}{4}\, \frac{\alpha_q q}{v_q}\, I_b(x + i 0)\,,
\label{eqn:P_b_II}
\end{equation}
where function $I_b(y)$ is given by Eqs.~\eqref{eqn:Ib}--\eqref{eqn:ImIb} of Sec.~\ref{sec:derivation}
and is computed by numerical quadrature.
The limiting forms of $P_b$ are as follows.
In the static limit we find
\begin{equation}
P_b(q,0)=-\frac{q}{4 v_q} C_b \alpha_q\,,
\end{equation}
where $C_b\approx0.48$.
In the optical limit, $\omega\gg v_q q$, we have
\begin{equation}
P_b(q,\omega) \simeq \frac{C'_b}{4}\, \frac{\alpha_q q^2}{i\omega}\,,
\quad
C'_b \approx -0.237\,,
\label{eqn:P_b_optical}
\end{equation}
which is consistent with\cite{Mishchenko2008mci} $C'_b = (8 - 3 \pi) / 6$.
Near the absorption threshold $|x - 1|\ll 1$,
we find the analytical expression
\begin{equation}
P_b(q,\omega) \simeq \frac{1}{6 \pi}\, \frac{\alpha_q q}{v_q}\,
\frac{1}{x - 1}\,
\left[\ln \left(\frac{8}{1 - x}\right) - 3\right],
\label{eqn:P_b_threshold}
\end{equation}
which agrees with the result of Ref.~\onlinecite{Gangadharaiah2008crf},
further providing the subleading divergent term (determined by the numerical factor inside the logarithm).

\section{Derivation}
\label{sec:derivation}

In this section we discuss how our results for the self-energy and vertex corrections have been derived. Within the Matsubara formalism~\cite{Mahan1990mpp}, the diagrams we compute are expressed by the integrals
\begin{widetext}
\begin{align}
P_a(q, i\Omega) &= -\frac{N}{\beta} \sum_{\nu} \int\frac{d^2{\bf k}}{(2\pi)^2}
{\rm tr}
[{\hat G}({\bf k},i\nu) {\hat G}({\bf k}+{\bf q},i\nu + i\Omega)
{\hat \Sigma}({\bf k}+{\bf q}) {\hat G}({\bf k}+{\bf q},i\nu + i\Omega)]\,,
\label{eqn:Pa}\\
P_b(q, i\Omega) &= -\frac{N}{\beta^2} \sum_{\nu,\nu'} \int\frac{d^2 {\bf k}}{(2\pi)^2} \frac{d^2 {\bf k}'}{(2\pi)^2}
V({\bf k}-{\bf k}')
{\rm tr}
[\hat{G}({\bf k},i\nu) \hat{G}({\bf k}+{\bf q},i\nu + i\Omega)
\hat{G}({\bf k}' + {\bf q}, i\nu' + i\Omega) \hat{G}({\bf k}',i\nu')]\,,
\label{eqn:Pb}
\end{align}
where $N = 4$ is the spectral degeneracy and the sums are performed over fermionic Matsubara frequencies $\nu = \pi (2 n + 1) \beta$.
Green's function $\hat{G}$, the self-energy matrix ${\hat \Sigma}$, and the Coulomb interaction kernel $V$ are given by
\begin{equation}
\hat{G}({\bf k}, i\nu) = (i\nu - v\,\mathbf{k} \cdot \bm{\sigma})^{-1}\,,
\quad
{\hat \Sigma}({\bf k}) = \frac{e^2}{4 \kappa} \mathbf{k}\cdot \bm{\sigma}
 \ln\frac{\Lambda}{|\mathbf{k}|}\,,
\quad
V(\mathbf{k})= \frac{2 \pi e^2}{\kappa |\mathbf{k}|}\,,
\label{eqn:V}
\end{equation}
where $\bm{\sigma} = \{\sigma_x, \sigma_y\}$ is the vector of the Pauli matrices.
To obtain results at real frequencies $\omega$
the analytical continuation $i\Omega \to \omega + i0$ has to be done at the end of the calculation, as usual.
In the zero temperature limit $\beta \to \infty$,
the integration over $\nu$ and the analytic continuation lead to
\begin{align}
2 P_a(q,\omega) &= 2 N\int\frac{d^2 \mathbf{k}}{(2\pi)^2}\,
\Sigma_k(1-\hat{n}_{\bf k} \cdot\hat{n}_{{\bf k}+{\bf q}})
\frac{E_{\mathbf{k},\mathbf{q}}^2 + \omega^2}
     {[E_{\mathbf{k},\mathbf{q}}^2 - (\omega + i0)^2]^2}\,,
\quad
E_{\mathbf{k},\mathbf{q}} \equiv v(|\mathbf{k}|+|\mathbf{k} + \mathbf{q}|)\,,
\quad
\hat{n}_{\mathbf{k}} \equiv \frac{\mathbf{k}}{|\mathbf{k}|}\,,
\label{eqn:Pa0}\\
P_b(q,\omega) &= -\frac{N}{2}\int\frac{d^2 \mathbf{k}}{(2\pi)^2}
\frac{d^2 \mathbf{k}'}{(2\pi)^2}
\frac{V(\mathbf{k} - \mathbf{k}')}{[E_{\mathbf{k},\mathbf{q}}^2 - (\omega + i0)^2]
                           [E_{\mathbf{k}',\mathbf{q}}^2 - (\omega + i0)^2]}
\Bigl\{
\omega^2 (\hat{n}_{\mathbf{k}} - \hat{n}_{\mathbf{k} + \mathbf{q}}) \cdot(\hat{n}_{\mathbf{k}'} - \hat{n}_{\mathbf{k}' + \mathbf{q}})
\notag\\
&+ E_{\mathbf{k},\mathbf{q}}E_{\mathbf{k}',\mathbf{q}}
[(1 - \hat{n}_{\mathbf{k}} \cdot \hat{n}_{\mathbf{k} + \mathbf{q}})
(1 - \hat{n}_{\mathbf{k}'} \cdot \hat{n}_{\mathbf{k}' + \mathbf{q}})
+ (\hat{n}_{\mathbf{k}} \times \hat{n}_{\mathbf{k} + \mathbf{q}}) \cdot
(\hat{n}_{\mathbf{k}'} \times \hat{n}_{\mathbf{k}' + \mathbf{q}})] 
\Bigr\}\,.
\label{eqn:Pb1}
\end{align}
These integrals can be simplified by transformation from the Cartesian $(k_x, k_y)$ to the elliptic coordinate system $(\mu, \nu)$, where $0 \leq \mu < \infty$ and $0 \leq \nu < 2\pi$.
In this system the coordinate grid is made of ellipses and hyperbolas
with the foci at ${\bf k}=0$ and ${\bf k}=-{\bf q}$.
The transformation formulas
are $k_x + i k_y = (q / 2) [\cosh(\mu + i\nu) - 1]$ and  $d^2 k = (q / 2)^2 (\cosh^2\mu - \cos\nu)$.
The integral for $2 P_a$ becomes
\begin{equation}\label{eqn:Pa_elliptic}
2 P_a(q,\omega)=\frac{N\alpha q}{8\pi^2 v}\int d\mu d\nu
\left[\ln \left(\frac{2\Lambda}{q}\right) -
      \ln (\cosh \mu - \cos \nu) \right]
(\cosh \mu-\cos \nu)
\frac{4\cosh^2 \mu + (2\omega / v q)^2}{[4\cosh^2 \mu - (2\omega / v q + i0)^2]^2}\,,
\end{equation}
which can be evaluated analytically in terms of the dilogarithm function $\mathrm{Li}_2(z)$.
For $\omega > 0$ we get Eq.~\eqref{eqn:Pa2} with
\begin{equation}
\begin{split}
I_a(x) &= \frac{1}{3}\frac{1+2x^2}{1-x^2} - \frac{x}{6}\frac{5-2 x^2}{1-x^2}
\ln \left(\frac{1-x}{1+x}\right)
-\frac{\pi}{12}\frac{3-12 \ln 2 +6x^2-4x^4}{(1-x^2)^{3/2}}\\
&-\frac{i}{(1-x^2)^{3/2}}
\left[ \frac{\pi^2}{4}-\mathrm{Li}_2\left(x + i\sqrt{1-x^2}\,\right)
 + \mathrm{Li}_2\left(-x-i\sqrt{1-x^2}\,\right)
 + \frac{i\pi}{2}\,\ln\left(x + i\sqrt{1-x^2}\,\right)\right]\,.
\end{split}
\label{eqn:I_a}
\end{equation}
The vertex correction to the polarizability is represented by the integral in Eq.~\eqref{eqn:Pb1}.
Although not amenable to analytic evaluation, this integral can still be simplified by employing the elliptic coordinates:
\begin{equation}
\begin{split}
P_b(q,\omega) &=-\frac{Nq\alpha}{32 \pi^3 v}\, \int\frac{d\mu d\mu' d\nu d\nu'}{\sqrt{\cosh(\mu+\mu')-\cos(\nu+\nu')} \sqrt{\cosh(\mu-\mu')-\cos(\nu-\nu')}}\,
\frac{\cosh\mu\cosh\mu'\sin\nu\sin\nu'}{(\cosh^2 \mu - y^2)(\cosh^2 \mu' - y^2)}\\
&\times
[(\sin\nu\sin\nu'+\sinh\mu\sinh\mu')+x^2(\cosh\mu\cosh\mu'+\tanh\mu\tanh\mu'\cos\nu\cos\nu')]\,,
\label{eqn:P_b_IV}
\end{split}
\end{equation}
where $y = x + i0$.
Integrating over $\nu$ and $\nu'$, one is lead to Eq.~\eqref{eqn:P_b_II} with $I_b$ given by the two-dimensional integral
\begin{gather}
I_b(y) = \frac{N}{\pi^3}\int_0^{\infty}\int_0^{\infty}
\frac{da db F(a, b)[\cosh a + (1 + 2 y^2) \cosh b]}
     {[1 - 2 y^2 + \cosh(a + b)] [1 - 2 y^2 + \cosh(a - b)]}\,,
\label{eqn:Ib}\\
F(a, b) = \frac{\cosh(a / 2)}{\cosh(b / 2)} \left\{ [K_b \cosh b - E_b (1 + \cosh b) ] E_a
- \frac{1}{3}\, [-E_a \cosh a + K_a(\cosh a - 1)] K_b \right\}\,,
\label{eqn:F}\\
K_\tau = K\left(\operatorname{sech}^2 \frac{\tau}{2}\right)\,,
\quad
E_\tau = E\left(\operatorname{sech}^2 \frac{\tau}{2}\right)\,,
\label{eqn:KE}
\end{gather}
\end{widetext}
where $K(z)$ and $E(z)$ denote the complete elliptic integrals of the first and the second kind.

\begin{figure}
\begin{center}
\includegraphics[width=2.5in]{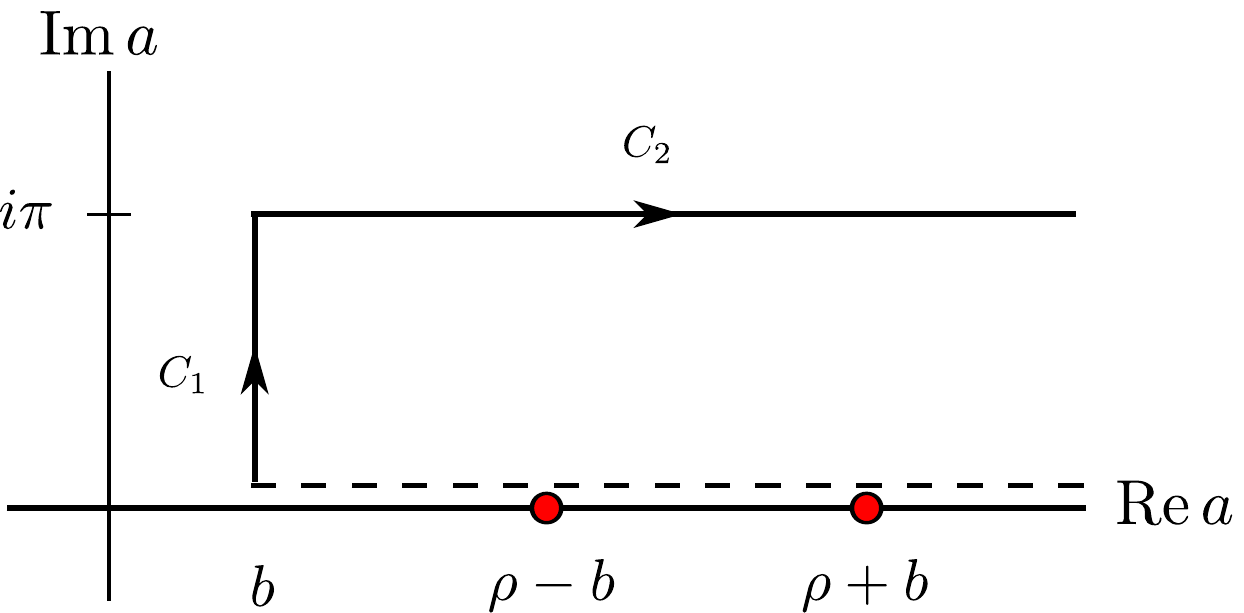}
\end{center}	
\caption{Integration contours in the complex $a$ plane to evaluate $\operatorname{Re} I_b(x)$ for $x>1$ in Eq.~\eqref{eqn:Ib}.
The dots indicate the poles of the integrand.
\label{fig:contour}
}
\end{figure}

For $x < 1$, the calculation of $P_b$ using these formulas is easily done numerically
and the result is real.
For $x > 1$, the standard numerical quadrature routines fail because the integration path in Eq.~\eqref{eqn:Ib} passes near the zeros of the denominator.
For such $x$, one can derive the following alternative formulas for $P_b$.
Denoting $x = \cosh(\rho/2)$, where $\rho > 0$ is real,
we see that the poles of the integrand are at $a_{1, 2} = \rho \mp b$.
The result for $I_b$ is complex, with the principal value integral (denoted by $\mathcal{P}$) giving the real part. The $a$ integral can be computed as the integral over contour depicted by the dashed line in Fig.~\ref{fig:contour}, which can be deformed into the union of contours $C_1$ and $C_2$. The real part of the integrand vanishes on $C_2$,
and thus only $C_1$ contributes to $\mathrm{Re}\, I_b$.

In turn, the imaginary part of $I_b$ can be obtained using the Sokhotski--Plemelj identity
\begin{equation}
\frac{1}{z \pm i 0} = \mathcal{P}\,\frac{1}{z} \mp i\pi \delta(z)\, ,
\label{eqn:Plemelji}
\end{equation}
followed by simple algebraic manipulations.
In the end, we obtain the alternative formulas for the real and imaginary parts of $I_b$ suitable for $x > 1$:
\begin{widetext}
\begin{align}
\mathrm{Re}\, I_b(x + i0) &= \int_0^\infty db\int_0^\pi \frac{du}{\cosh \rho -\cos u}
\mathrm{Im}\, \left[\frac{F_s(b + iu, b)}{\cosh(2b + iu)-\cosh\rho}\right]\,,
\label{eqn:reIb}\\
\mathrm{Im}\, I_b(x + i0) &= \frac{1}{\pi^2 \sinh \rho \sinh(\rho/2)}
\int_0^\infty \frac{d\mu}{\sinh \mu}
\left[\frac{F_s(\mu + \rho,\rho)}{\cosh(\mu + \rho/2)} - \frac{F_s(\mu - \rho,\rho)}{\cosh(\mu - \rho/2)}\right]\,,
\label{eqn:ImIb}\\
F_s(a, b) &= F(a, b) [\cosh a +(\cosh \rho + 2) \cosh b]+ (a \leftrightarrow b)\,,
\quad
\rho = 2 \ln \left(x + \sqrt{x^2 - 1}\,\right)\,.
\end{align}
\end{widetext}
Straightforward numerical evaluation and asymptotic analysis of these expressions lead
to the results presented in Sec.~\ref{sec:results} above.

Since all the above calculations have been done within the Dirac approximation,
it is instructive to discuss how they can be generalized to a more realistic lattice model.
We will consider in particular the self-energy term (the vertex correction can be analyzed similarly).
We will show that the correction calculated from the Dirac model, Eq.~\eqref{eqn:Pa0},
and that computed from the lattice model is vanishingly small in the limit of interest,
$\{\omega \ll v / |\bm{a}|, q \ll 1/ |\bm{a}|\}$,
where $\bm{a}$ is the vector connecting a pair of nearest lattice sites.

Let the kinetic energy matrix on the lattice be
$\hat{H} = \mathcal{H}_{\mathbf{k}}\cdot \bm{\sigma}$, where $\mathcal{H}_{\mathbf{k}}=(\mathcal{H}^x_{\mathbf{k}},\mathcal{H}^y_{\mathbf{k}})$, and
let the on-site density distributions be described by a form-factor
function $\mathcal{F}(\mathbf{q})$,
which is close to unity at $q \ll 1 / b$ and rapidly decays to zero at $q \gg 1 / b$,
where $b \sim a$ is the characteristic size of the site orbitals.
In terms of these notations
the self-energy on the lattice
$\hat{\Sigma}_{\mathbf{q}}=\mathbf{\Sigma}_{\mathbf{q}} \cdot \bm{\sigma}$, with $\mathbf{\Sigma}_{\mathbf{q}}=(\Sigma^x_{\mathbf{q}},\Sigma^y_{\mathbf{q}})$,
is given by the equation
\begin{widetext}
\begin{equation}
\Sigma^x_\mathbf{q} + i \Sigma^y_\mathbf{q}
= \int\frac{d^2 \mathbf{k}}{(2\pi)^2}
\mathcal{V}({\bf q}-{\bf k}) e^{i(\mathbf{q} - \mathbf{k}) \cdot \bm{a}} (\hat{\mathcal{H}}^x_{\mathbf{k}} + i \hat{\mathcal{H}}^y_{\mathbf{k}}) \,,
\quad \mathcal{V}(\mathbf{q})
=\frac{2 \pi e^2}{\kappa |\mathbf{q}|}
\mathcal{F}(\mathbf{q})\,,\quad
\hat{\mathcal{H}}_{\mathbf{k}} \equiv \frac{\mathcal{H}_\mathbf{k}}{|\mathcal{H}_\mathbf{k}|} \,.
\label{eqn:Sigma_lat}
\end{equation}
Note that the form-factor $\mathcal{F}(\mathbf{q})$ regularizes the \emph{short-range}
behavior of the interaction potential
and serves as a cutoff on the momentum transfer introduced in
Refs.~\onlinecite{Mishchenko2008mci, Sheehy2009oto}.
Note also that the integration in Eq.~\eqref{eqn:Sigma_lat} is over the entire momentum space to account for Umklapp processes.

From Eq.~\eqref{eqn:Sigma_lat} we see that the self-energy in the lattice model and the Dirac model [Eq.~\eqref{eqn:Sigma_q}] have the same functional form if the deviation $\delta \mathbf{k}$ of $\mathbf{k}$ from a corner of the BZ $\mathbf{K}$ is small:
\begin{equation}
\mathbf{\Sigma}_{\mathbf{K}+\delta\mathbf{k}}
= \frac{e^2}{4\kappa}\, \delta\mathbf{k}
\ln \left|\frac{\Lambda_{\text{lat}}}{\delta \mathbf{k}}\right|
\left[1 + \mathcal{O}\left(\left|\frac{\delta \mathbf{k}}{\Lambda_{\text{lat}}} \right|\right)\right]\,,
\quad
\Lambda_{\text{lat}} \sim \frac{1}{a}\,.
\end{equation}
The lattice analog of Eq.~\eqref{eqn:Pa0} is
\begin{equation}
2 P^{\mathrm{lat}}_a(q,\omega)
= N_s \int_{\mathrm{BZ}}\frac{d^2 \mathbf{k}}{(2\pi)^2} \mathbf{\Sigma}_{\mathbf{k}}\cdot(\hat{\mathcal{H}}_{\mathbf{k}}
- \hat{\mathcal{H}}_{\mathbf{k} + \mathbf{q}}) 
\frac{\mathcal{E}_{\mathbf{k},\mathbf{q}}^2 + \omega^2}
     {[\mathcal{E}_{\mathbf{k},\mathbf{q}}^2 - (\omega + i0)^2]^2}\,,
\quad
\mathcal{E}_{\mathbf{k},\mathbf{q}} \equiv |\mathcal{H}_\mathbf{k}|+|\mathcal{H}_{\mathbf{k} + \mathbf{q}}|\,,
\label{eqn:P_a_lat}
\end{equation}
\end{widetext}
where the integral is now taken over a Brillouin zone (BZ) and $N_s = 2$ accounts for the spin degeneracy only.
It is easy to see that this integral is convergent
and that the difference between Eq.~\eqref{eqn:Pa0} and Eq.~\ref{eqn:P_a_lat} is of the order of $(q / \Lambda_{\text{lat}}) \ln (q / \Lambda_{\text{lat}})$.
This difference is vanishingly small in the limit of interest, as we stated above.

\section{Discussion}
\label{sec:discussion}

In this section we compare our work with previous literature and discuss some effects beyond the first-order perturbation theory.

The interaction correction to the static dielectric function $\epsilon(q, 0)$ has been considered in Ref.~\onlinecite{Kotov2008eei}.
Our Eq.~\eqref{eqn:epsilon_dc} for this quantity differs from the result obtained therein by the numerical coefficient of the quadratic term,
$0.778$ versus $0.53$.
This discrepancy is for two reasons.
First is the apparent error in the numerical evaluation of the vertex diagram in Ref.~\onlinecite{Kotov2008eei}.
Second is the different treatment of the self-energy contribution.
In Ref.~\onlinecite{Kotov2008eei}, this contribution is assumed to be completely absorbed into the velocity renormalization.
In our renormalization scheme, only the first term in Eq.~\eqref{eqn:Pa2} is absorbed while the second leaves a finite remainder, Eq.~\eqref{eqn:Pa_dc}.
This difference is not merely a matter of convention.
It has to do with the principal distinction regarding the relations between observable and nonobservable quantities.
While the former relations do not depend on the renormalization scheme, the latter do.
In our case,
the expansion of the static dielectric function $\epsilon(q, 0)$ in powers of $\alpha_q$ (defined by the quasiparticle phase velocity $v_q$) is unique and given by Eq.~\eqref{eqn:epsilon_dc}.
On the other hand, the expansion of $\epsilon(q, 0)$ in powers of the ``bare'' coupling $\alpha$ has the form
\begin{equation}
\epsilon(q, 0) \simeq 1 + \frac{\pi}{2}\, \alpha + \left(0.778 - \frac{\pi}{8}\, \ln \frac{\Lambda}{q}\right) \alpha^2\,,
\label{eqn:epsilon_dc_II}
\end{equation}
in which the coefficient for $\alpha^2$ depends on the nonuniversal cutoff parameter $\Lambda$.
Clearly, this coefficient can be discussed only after the renormalization scheme is precisely defined,
as we have done here.

Next, the optical limit of the polarization function has been vigorously debated~\cite{Mishchenko2008mci, Sheehy2009oto, Katsnelson2008opo, Juricic2010coi, Giuliani2012erg} in the context of the interaction correction to the universal conductivity of graphene. Our method of calculation and the final result, Eq.~\eqref{eqn:sigma_I},
are the same as in Ref.~\onlinecite{Mishchenko2008mci}.
The interaction correction to the optical conductivity of \emph{doped} graphene
has been studied in Ref.~\onlinecite{Abedinpour2011dwp}.
In the limit $\omega \gg \mu$,
where $\mu$ is the chemical potential measured with respect to the Dirac point,
one expects to recover the behavior characteristic of neutral graphene.
In this limit the results of Ref.~\onlinecite{Abedinpour2011dwp} agree with those of Ref.~\onlinecite{Mishchenko2008mci} and therefore with ours.
The origin of the discrepancy with Ref.~\onlinecite{Juricic2010coi}
is discussed in Appendix~\ref{sec:optical}.

\begin{figure}
\begin{center}
\includegraphics[width=2.8in]{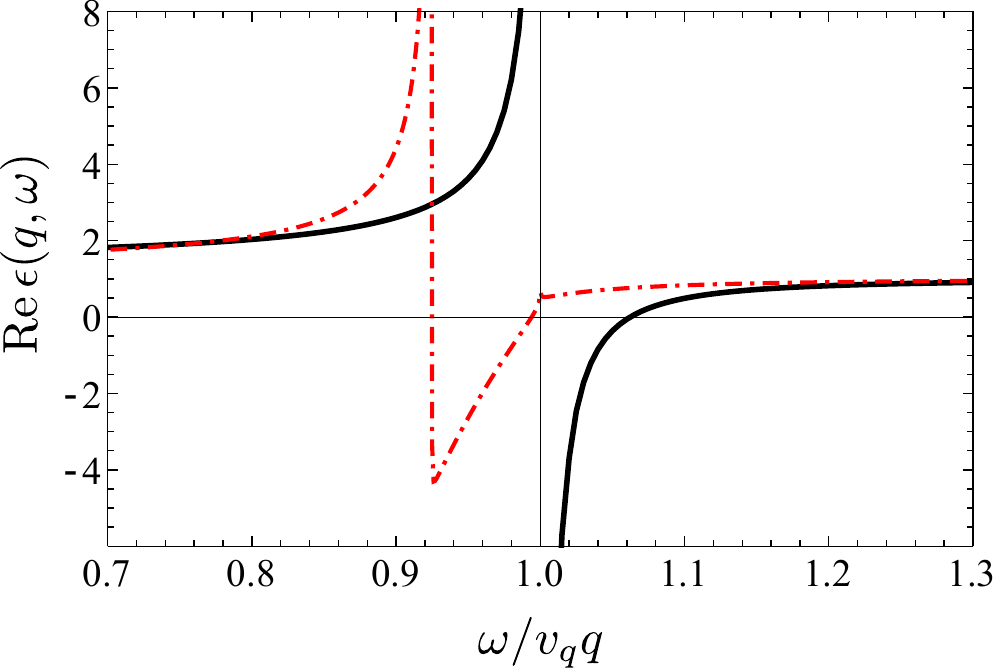}
\end{center}	
\caption{(Color online)
The real part of the dielectric function near the absorption threshold for $\alpha_q = 0.3$.
The black solid line is our first-order theory,
the red dash-dotted line is the ladder sum from Ref.~\onlinecite{Gangadharaiah2008crf}.
\label{fig:threshold}
}
\end{figure}

Finally, the behavior of $P(q, \omega)$ near the absorption threshold $|\omega/{v_q q} - 1|\ll 1$
has been studied by the authors of Ref.~\onlinecite{Gangadharaiah2008crf} (henceforth GFM).
In this special region the perturbative expansion of the dielectric function diverges, see Fig.~\ref{fig:threshold}.
The divergence of the self-energy term is stronger, $\sim (1 - x^2)^{-3 / 2}$ [Eq.~\eqref{eqn:Pa_theshold}] than that $\sim (1 - x)^{-1} \ln (1 - x)$ [Eq.~\eqref{eqn:P_b_threshold}] of the vertex term.
However, GFM argued that the divergence of the self-energy term can be trivially absorbed into the velocity renormalization thus leaving only the divergence of the vertex.
Summing the ladder series for the vertex corrections,
they obtained a nonperturbative expression for $\epsilon(q, \omega)$.
This expression is also plotted in Fig.~\ref{fig:threshold} assuming the renormalized velocity of GFM coincides with $v_q$.

An important qualitative prediction of GFM theory is vanishing of $\mathrm{Re}\, \epsilon$ at certain $x < 1$,
see Fig.~\ref{fig:threshold}, which signals the presence of a new collective mode --- ``excitonic plasmon.''
While we find this prediction very interesting,
we wish to express some reservations in the validity of GFM approach.
We believe that the nonperturbative treatment should begin with the resummation of the self-energy not the vertex term because the former is more divergent.
Such a resummation would make the quasiparticle velocity $v_q$ momentum-dependent, cf.~Eq.~\eqref{eqn:vel_I}.
In other words, the linear Dirac spectrum would be replaced by a spectrum with a finite curvature.
As in the case of the static response, it is not possible to faithfully represent the effect of such a curvature
by simply replacing $v$ with another constant number.
It is easy to see that the
%
%
finite curvature of the spectrum modifies the behavior of the dielectric function over the range of frequencies $\sim \alpha q$,
which is much wider than the interval $\alpha^2 q$ where the higher-order terms considered by GFM are important,
see also Fig.~\ref{fig:threshold}.
Our preliminary analysis suggests that this significantly modifies the analytical structure of the ladder sum compared to what was obtained by GFM.
This intriguing problem warrants further study.

Our work is supported by the Grants NSF PHY11-25915, by UCOP (M.M.F.), by the KITP Graduate Fellows Program (I.S.), the Welch Foundation Grant No. TBF1473 (I.S.) and the NRI SWAN program (I.S.). We are grateful to the KITP at UCSB, where this work has been carried out, for hospitality. We are thankful to I.~Herbut, P.~Koroteev, V.~Kotov, V.~Mastropietro, E.~Mishchenko, M.~Polini, O.~Vafek, and B.~Uchoa for discussions and comments on the manuscript. We thank M.~Vozmediano for bringing Ref.~\onlinecite{Teber} to our attention.

\appendix
\section{Optical limit}
\label{sec:optical}

In this Appendix we review the derivation of the polarization function in the limit $v_q q \ll \omega$
and attempt to settle the existing dispute in the literature. In this limit the requisite integrals simplify so that
they can be evaluated analytically.
Equation~\eqref{eqn:Pa0} for the self-energy term can be expanded to second order in $q$
and reduces to
\begin{equation}
2 P_a(q, \omega) \simeq \int \frac{d^2 k}{(2\pi)^2}\,
\frac{2 q^2 (\omega^2 + 4 v^2 k^2)\, \Sigma_{k}}
     {k^2 [4 v^2 k^2 - (\omega + i0)^2]^2}\,.
\label{eqn:P_a_II}
\end{equation}
The integration is convergent and yields\cite{Mishchenko2008mci}
\begin{equation}
2 P_a(q,\omega) \simeq \frac{1}{16}\,
\frac{\alpha q^2}{i \omega}
\label{eqn:P_a_III}
\end{equation}
independently of $\Lambda$, which is our Eq.~\eqref{eqn:P02Pa_optical}.
In turn, the integral in Eq.~\eqref{eqn:Pb1} for the vertex correction
is also convergent and has the value of\cite{Mishchenko2008mci}
\begin{equation}
P_b(q, \omega) \simeq \frac{8-3\pi}{24}\, \frac{\alpha q^2}{i \omega}\,,
\label{eqn:P_b_III}
\end{equation}
in agreement with our Eq.~\eqref{eqn:P_b_optical}.
Combining Eqs.~\eqref{eqn:sigma}, \eqref{eqn:P_a_III}, and \eqref{eqn:P_b_III},
one arrives at Eq.~\eqref{eqn:sigma_I}.

This straightforward calculation was questioned in Ref.~\onlinecite{Juricic2010coi} (henceforth referred to as JVH) where the twice larger value of $2 P_a$ was obtained.
JVH used the dimensional regularization (DR) scheme in which the fermions are initially assumed to live in $D = 2 - \varepsilon$ dimensions, with the limit $\varepsilon \to 0$ taken at the end of the calculation. Let us critically analyze this approach.

Correcting several innocuous mistakes of JVH [such as
the overall sign error in their Eqs.~(62) and its carry over into Eqs.~(F3)--(F12); also,
the missing factor of $4$ in front of $k^2$ in the denominator of Eq.~(F7)], combining together
their Eqs.~(F7) and (F9),
doing the analytic continuation from imaginary to real frequency,
and finally, temporarily setting $v \equiv 1$,
we see that the DR approach predicts
\begin{equation}
\begin{split}
2 P_a^{\mathrm{DR}}(q, \omega) &= (D - 1) \int \frac{d^D k}{(2\pi)^D}\,
 \frac{2 q^2 (\omega^2 + 4 k^2)\, \Sigma_k^{\mathrm{DR}}}
      {k^2 [4 k^2 - (\omega + i0)^2]^2}\\
&\simeq
2 q^2 \int_0^\infty d k \nu_\varepsilon(k)
\frac{(\omega^2 + 4 k^2)\, \Sigma_k^{\mathrm{DR}}}{k^2 [4 k^2 - (\omega + i0)^2]^2}
\,.
\end{split}
\label{eqn:P_a_DR_I}
\end{equation}
Here function $\nu_\varepsilon(k)$ given by
\begin{equation}
\nu_\varepsilon(k) = \frac{4}{(4\pi)^{D/2} \Gamma(D/2)}\, |k|^{D - 1}
\label{eqn:nu_D_I}
\end{equation}
generalizes the free-fermion density of states (DOS) $\nu_0(k)$ to the case of $D$ dimensions
and $\Gamma(z)$ is the Euler Gamma-function.
The self-energy $\Sigma_k^{\mathrm{DR}}$ in Eq.~\eqref{eqn:P_a_DR_I} is given by
\begin{equation}
\Sigma_k^{\mathrm{DR}} = \frac{e^2}{4}\, \lambda^\varepsilon\,
\frac{\Gamma\left(1-\frac{D}{2}\right)
      \Gamma\left(\frac{D + 1}{2}\right)
      \Gamma\left(\frac{D - 1}{2}\right)}{
      (4\pi)^{D/2} \Gamma(D)}\, k^{D - 1}\,,
\label{eqn:Sigma_DR_I}
\end{equation}
where we introduced the extra factor $\lambda^\varepsilon$ compared to Eq.~(65) of JVH, as it is commonly done.
This factor makes the (energy) units correct and goes away at the end of the calculation.
However, it is convenient to think that $\lambda$ is of the order of 
$\omega / v$, which is the typical momentum of electrons and holes that determine the response at frequency $\omega$.

The evaluation of the integral in Eq.~\eqref{eqn:P_a_DR_I} indeed gives
the twice larger value than Eq.~\eqref{eqn:P_a_III} in the limit $\varepsilon \to 0$.
Hence, the discrepancy in hand is not due to a mathematical mistake of JVH.

The debate held by the authors of Refs.~\onlinecite{Mishchenko2008mci, Sheehy2009oto} on the one side and JVH on the other revolved around the question of whether the ultraviolet (UV) regularization is done in a physically correct manner.
Examples of the regularization schemes that have been discussed include the cutoff on the interaction potential,
$V({\mathbf{k}}) \to V({\mathbf{k}}) \exp(-k / \Lambda)$ and the cutoff on the bandwidth,
$\nu_0(k) \to \nu_0(k) \Theta(\Lambda - k)$.
In the DR scheme, the interaction potential $V({\mathbf{k}})$ is unchanged but the DOS is modified:
\begin{equation}
\nu_\varepsilon(k) \simeq  
\frac{1 +\mathcal{O}(\varepsilon)}{\lambda^\varepsilon}\, \nu_0(k)
\left|\frac{\lambda}{k}\right|^\varepsilon\,.
\label{eqn:nu_D_II}
\end{equation}
Note that both $\nu_0(k)$ and $\nu_\varepsilon(k)$ are the \emph{bare} DOS not the renormalized one, which would follow the spectrum described by Eq.~\eqref{eqn:vel_I}.

The soft UV cutoff is assured by the last factor in Eq.~\eqref{eqn:nu_D_II}.
It is nearly unity at the physically relevant momenta $k \sim \lambda$ but gradually decreases with $k$ thereby suppressing the bare DOS below that of 2D Dirac fermions $\nu_0(k)$.
The characteristic momentum cutoff scale is $\Lambda \sim \lambda e^{1/\varepsilon}$ at which the suppression factor is equal to $1 / e$.
The same conclusion follows from the comparison of the self-energy in the DR scheme, Eq.~\eqref{eqn:Sigma_DR_I},
\begin{equation}
\Sigma_k^{\mathrm{DR}} =
      \frac{e^2}{4}\, k \left[\frac{1}{\varepsilon} + \ln \frac{\lambda}{k} + \mathcal{O}(1)\right]\,,
\label{eqn:Sigma_DR_II}
\end{equation}
with Eq.~\eqref{eqn:Sigma_q}.
Accordingly,
the comparison between the DR and other cutoff schemes can be done by keeping $\varepsilon$ small but finite:
\begin{equation}
\frac{1}{\varepsilon} \equiv \mathcal{L}
 = \ln \left(\frac{\Lambda}{\lambda}\right) + \mathcal{O}(1) \gg 1\,.
\label{eqn:L}
\end{equation}
This kind of interpretation of the parameter $1 / \varepsilon$ is well known in the practice of DR.
The problem with the DR approach is therefore not the implementation of the UV cutoff but the behavior at small $k$.
Within the DR scheme the correction to the bare DOS is divergent at small $k$:
\begin{equation}
\nu_\varepsilon(k) \simeq  
\frac{\nu_0(k)}{\lambda^\varepsilon}\, 
\left(1 + \frac{1}{\mathcal{L}}\, \ln \frac{\lambda}{k}\right)\,.
\label{eqn:nu_D_III}
\end{equation}
Substituting Eqs.~\eqref{eqn:Sigma_DR_II}--\eqref{eqn:nu_D_III} into Eq.~\eqref{eqn:P_a_DR_I},
we get
\begin{widetext}
\begin{equation}
2 P_a^{\mathrm{DR}}(q, \omega) \simeq
2 q^2 \int_0^\infty d k \nu_0(k) \left(1 + \frac{1}{\mathcal{L}}\, \ln \frac{\lambda}{k}\right)  \frac{e^2}{4}\, k \left(\mathcal{L} + \ln \frac{\lambda}{k}\right)
\frac{\omega^2 + 4 k^2}{k^2 [4 k^2 - (\omega + i0)^2]^2}
\simeq \frac{e^2}{8}\, \frac{q^2}{i \omega}\,.
\label{eqn:P_a_DR_II}
\end{equation}
In comparison, Eq.~\eqref{eqn:P_a_II} written in the same units convention reads
\begin{equation}
2 P_a(q, \omega) = 2 q^2 \int_0^\infty d k \nu_0(k)
\frac{e^2}{4}\, k \left(\mathcal{L} + \ln \frac{\lambda}{k}\right)
\frac{\omega^2 + 4 k^2}{k^2 [4 k^2 - (\omega + i0)^2]^2}
= \frac{e^2}{16}\, \frac{q^2}{i \omega}\,.
\label{eqn:P_a_IV}
\end{equation}
\end{widetext}

The only difference between the two expressions is in the bare DOS.
The extra contribution to the integral in Eq.~\eqref{eqn:P_a_DR_II} compared to that in Eq.~\eqref{eqn:P_a_IV} comes from the product of the artificial
$\nu_\varepsilon(k) - \nu_0(k) \propto 1 / \mathcal{L}$ correction to the bare DOS and the logarithmically large $\Sigma_k^{\mathrm{DR}} \propto  \mathcal{L}$ self-energy\cite{Note2}.
Since our Eq.~\eqref{eqn:P_a_IV} correctly describes the physically motivated lattice model (see Sec.~\ref{sec:derivation}) while the DR scheme of JHV gives a different result,
in our opinion the latter does not faithfully represent the behavior of electrons in graphene.

In closing,
we would like to comment on the
ultra-relativistic regime of extremely low frequencies and momenta at which the renormalized velocity $v_q$ approaches the speed of light $c$ while
$\alpha_q$ approaches the fine-structure constant $\alpha_0 = e^2 / \hbar c \approx 1 / 137$. This limit is of no practical importance for real graphene samples.
According to Eq.~\eqref{eqn:alpha_q},
it is reached at unobservably small frequencies and momenta of the order of
\begin{equation}
\Lambda_{\text{rel}} \equiv \Lambda e^{-4 / \alpha_0}\,.
\label{eqn:Lambda_rel}
\end{equation}
Nevertheless, it is an interesting limit from a formal standpoint because it represents a fixed point of the renormalization group~\cite{Gonzalez1994nfl, Vozmediano2011}.

Interaction corrections to the conductivity in the relativistic regime have been found to be~\cite{Teber}
\begin{equation}
\frac{\sigma(0, \omega)}{\sigma_0} - 1
\simeq \frac{92 - 9\pi^2}{18 \pi}\, \alpha_0\,.
\label{eqn:sigma_Teber}
\end{equation}
The smallness of the numerical coefficient $(92 - 9\pi^2) / (18 \pi) \approx 0.056$,
together with the smallness of its nonrelativistic counterpart [Eq.~\eqref{eqn:sigma_I}]
suggests that such coefficient remains small throughout the crossover from the nonrelativistic ($\omega \gg \Lambda_{\text{rel}}$) to relativistic ($\omega \ll \Lambda_{\text{rel}}$) regime.
It is interesting that Ref.~\onlinecite{Teber} also employed a DR scheme.
Unlike in JVH treatment, the low-energy electron DOS has been preserved because the DR was performed on the $3 +(1 -\varepsilon)$ dimensions of the $U(1)$ photon field while the electrons remain in $2+1$ dimensions.

Another recent work~\cite{Giuliani2012erg} has presented results for the frequency-dependent optical conductivity in the relativistic regime at the one-loop level of the exact
renormalization group on the same footing as those found in the nonrelativistic model~\cite{Mishchenko2008mci, Sheehy2009oto, Katsnelson2008opo, Juricic2010coi} we studied here.
We think this is misleading for two reasons.
First, the two results are expected to apply at very different frequency scales,
$\omega \ll \Lambda_{\text{rel}}$ and $\omega \gg \Lambda_{\text{rel}}$.
Their comparison is meaningful only at the
common border of validity $\omega \sim \Lambda_{\text{rel}}$ and only
by the order of magnitude.
Second, it seems that such a comparison would require a two-loop calculation, as in Ref.~\onlinecite{Teber}.

%

\end{document}